\documentclass[12pt,preprint]{aastex}

\begin{document}

\title{An Investigation of the Absolute Proper Motions of the SCUSS Catalog}

\author{Xiyan Peng\altaffilmark{1,2}, Zhaoxiang Qi\altaffilmark{3}, Zhenyu Wu\altaffilmark{1,\star}, Jun
Ma\altaffilmark{1}, Cuihua Du\altaffilmark{2}, Xu
Zhou\altaffilmark{1}, Yong Yu\altaffilmark{3}, Zhenghong
Tang\altaffilmark{3}, Zhaoji Jiang\altaffilmark{1}, Hu
Zou\altaffilmark{1}, Zhou Fan\altaffilmark{1}, Xiaohui
Fan\altaffilmark{4}, Martin C. Smith\altaffilmark{3}, Linhua
Jiang\altaffilmark{5}, Yipeng Jing\altaffilmark{6}, Mario G.
Lattanzi\altaffilmark{7}, Brian J. McLean\altaffilmark{8}, Michael
Lesser\altaffilmark{4}, Jundan Nie\altaffilmark{1}, Shiyin
Shen\altaffilmark{3}, Jiali Wang\altaffilmark{1}, Tianmeng
Zhang\altaffilmark{1}, Zhimin Zhou\altaffilmark{1}, Songhu
Wang\altaffilmark{9}}

\email{zywu@nao.cas.cn} \altaffiltext{1}{Key Laboratory of Optical
Astronomy, National Astronomical Observatories, Chinese Academy of
Sciences, Beijing, 100012, China}

\altaffiltext{2}{University of Chinese Academy of Sciences, Beijing,
100039, China}

\altaffiltext{3}{Shanghai Astronomical Observatory, Chinese Academy
of Sciences, Shanghai 200030, China }

\altaffiltext{4}{Department of Astronomy and Steward Observatory,
University of Arizona, Tucson, Arizona, USA}

\altaffiltext{5}{Kavli Institute for Astronomy and Astrophysics,
Peking University, Beijing 100871, China}

\altaffiltext{6}{Department of Physics and Astronomy, Shanghai Jiao
Tong University, Shanghai 200240}

\altaffiltext{7}{INAF--Osservatorio Astrofisico di Torino, Strada
Osservatorio 20, 10025 Pino Torinese, TO, Italy.}

\altaffiltext{8}{Space Telescope Science Institute, 3700 San Martin
Drive,  Baltimore, MD 21218, USA.}

\altaffiltext{9}{School of Astronomy and Space Science and Key
Laboratory of Modern Astronomy and Astrophysics in Ministry of
Education, Nanjing University, Nanjing 210093}

\begin{abstract}
Absolute proper motions for $\sim$ 7.7 million objects were derived
based on data from the South Galactic Cap $u$-band Sky Survey
(SCUSS) and astrometric data derived from uncompressed Digitized Sky
Surveys that the Space Telescope Science Institute (STScI) created
from the Palomar and UK Schmidt survey plates. We put a great deal
of effort into correcting the position-, magnitude- and color-
dependent systematic errors in the derived absolute proper motions.
The spectroscopically confirmed quasars were used to test the
internal systematic and random error of the proper motions. The
systematic errors of the overall proper motions in the SCUSS catalog
are estimated as $-0.08$ and $-0.06\,\rm{mas/yr}$ for
$\mu_{\alpha}\rm{cos}\delta$ and $\mu_{\delta}$, respectively. The
random errors of the proper motions in the SCUSS catalog are
estimated independently as 4.2 and $4.4 \,\rm{mas/yr}$ for
$\mu_{\alpha}\rm{cos}\delta$ and $\mu_{\delta}$. There is no obvious
position-, magnitude- and color- dependent systematic errors of the
SCUSS  proper motions. The random error of the proper motions goes
up with the magnitude from about $3 \,\rm{mas/yr}$ at $u =
18.0\,\rm{mag}$ to about $7 \,\rm{mas/yr}$ at $u = 22.0\,\rm{mag}$.
The proper motions of stars in SCUSS catalog are compared with those
in the SDSS catalog, and they are consistent well.
\end{abstract}


\section{INTRODUCTION}
A catalog of proper motions in the {\bf International} Celestial
Reference System (ICRS), with precise positions and proper motions
as well as deep {\bf limiting} magnitude, is important for {\bf
research} on the Galactic structure, Galactic kinematics and stellar
{\bf populations}.

There are two methods to determine the absolute proper motions of
stars based on the hypothesis that the proper motions of background
galaxies are zero. One is that all the observations obtained in
different epochs are transformed into one reference position system
using galaxies, and absolute proper motions are calculated directly.
For example, Munn et al. (2004) took this method to construct the
improved SDSS proper motion catalog combining USNO-B (Monet et al.
2003) and SDSS (Gunn et al. 1998; York et al. 2000; Lupton et al.
2001; Stoughton et al. 2002). The basic procedure in this work is as
follows. For each object from USNO-B, the nearest 100 galaxies
(classified as $galaxy$ by SDSS morphological
classification\footnote{http://www.sdss.org/dr1/products/catalogs/flags.html})
detected in both SDSS and USNO-B were selected. Then the mean
offsets in right ascension and declination between the SDSS and
USNO-B positions were measured from the 100 nearest galaxies, and
were added to the USNO-B position of this object. Finally the proper
motions of objects in SDSS were computed using a linear fit with the
re-calibrated USNO-B positions and the SDSS positions at different
epochs.

The other method to determine absolute proper motions is to take a
novel calibration method which iteratively uses the stellar sources
and galaxies on each position system to eliminate all systematic
errors between different position systems. Absolute Proper motions
Out side the Plane (APOP) adopted this technique with the Guide Star
Catalog \uppercase\expandafter{\romannumeral2}
(GSC-\uppercase\expandafter{\romannumeral2}) Schmidt plate data (Qi
et al. 2012; Lasker et al. 2008). Meanwhile, the APOP used the
measured coordinates ($x$, $y$) of objects instead of equatorial
coordinates to calculate the absolute proper motions. The principal
calibration steps for APOP are as follows.
\begin{enumerate}
\item Removing the position-dependent systematic errors (PdE) with a moving-mean filter using good-quality stellar objects;
\item Selecting galaxies from non-point-like sources (non-stars) via their common null-motion characteristics;
\item Calibrating magnitude- and color- dependent (MdE and CdE) errors and the residual PdE of all objects with respect to the galaxies selected above;
\item Calculating absolute proper motions from all-epoch plate data with a linear fit.
\end{enumerate}

The construction of the latest version of APOP catalog and
evaluation of the catalog are described in detail by Qi et al.
(2014). In this paper, we take almost the same method and pipeline
as APOP to derive absolute proper motions.

The SCUSS (Zou et al. 2014; Zhou et al. 2014; Jia et al. 2014) is a
$u$-band (3538 \AA) imaging survey using the $2.3\,\rm{m}$ Bok
telescope. The survey covers about 5000 square degrees. Most of the
SCUSS fields were observed twice in the course of survey from 2010
to 2013. It is difficult to measure accurate proper motions for the
SCUSS objects from the SCUSS data only. In this paper, we construct
a SCUSS proper motion catalog by combining the SCUSS data with the
GSC-\uppercase\expandafter{\romannumeral2} Schmidt plate data. The
GSC-\uppercase\expandafter{\romannumeral2} Schmidt plate data is an
all-sky astro-photometric data base which were derived from the
digitization of the Palomar and UK Schmidt surveys. The
GSC-\uppercase\expandafter{\romannumeral2} contains astrometry,
photometry and classification for objects down to the limiting
magnitude of the plates .

This paper is organized as follows. In Section 2, we summarize the
SCUSS program and GSC-\uppercase\expandafter{\romannumeral2} project
that are used to derive the SCUSS proper motion catalog. Section 3
describes the construction of the SCUSS proper motion catalog. In
Section 4, quasars are used to perform internal test of the proper
motions. We compare the catalog with the SDSS proper motion catalog
in Section 5. We summarize the paper in Section 6.

\section{DATA}

\subsection{The SCUSS Program}

The SCUSS is a $u$-band imaging survey with an effective wavelength
of $\sim$ 3538 \AA~and a bandwidth $\sim$ 345 \AA~. This $u$-band
filter is slightly bluer than the SDSS $u$-band filter. The imaging
depth is $\sim$ $23.5\,\rm{mag}$, roughly $1.5\,\rm{mag}$ deeper
than that of the SDSS imaging survey. The SCUSS project began its
observations in September 2010 and ended in October 2013. It covers
a total 5000 square degrees. Proper motions are measured with
respect to a reference frame of galaxies. Hence, the accuracy of
star/galaxy classification affects systematic error of the proper
motions. The SCUSS proper motion catalog covers 3700 square degrees
(Galactic latitude $b < -30^\circ$ and {\bf Equatorial latitude
$\delta> -10^\circ$}). In this area, the accuracy of star/galaxy
classification is reliable. The spatial distribution of the SCUSS
proper motion catalog is shown in Figure 1. Most of the images in
this area were taken with seeing $<3''$, sky background $< 500$ ADU.

\begin{figure}[!htb]
\figurenum{1} \center \resizebox{\width}{!}{\rotatebox{0}
{\includegraphics[width=0.9\textwidth]{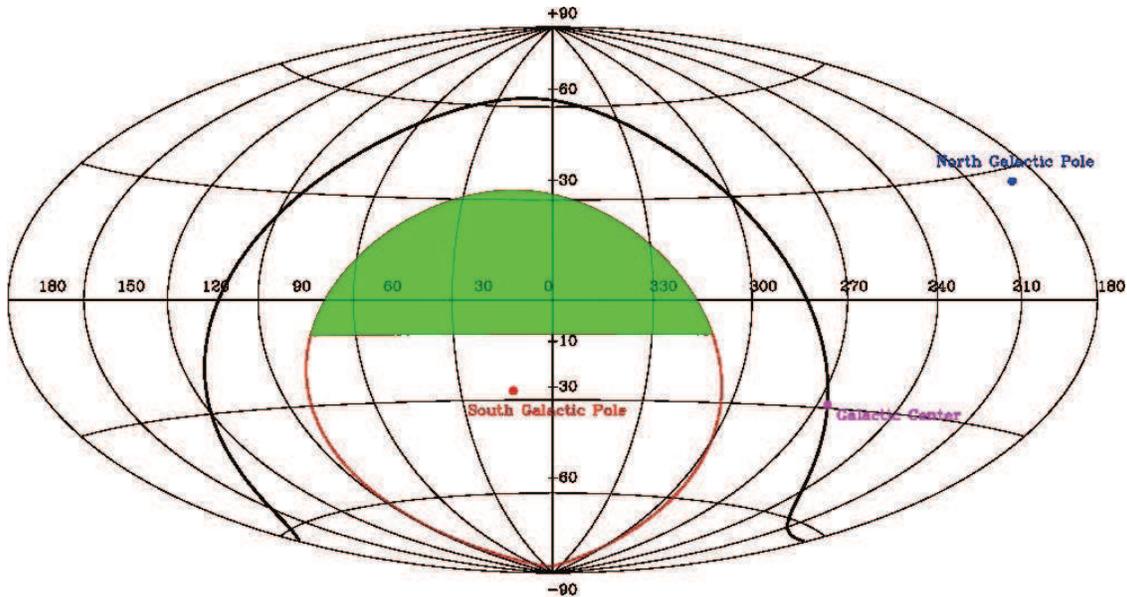}}} \caption{Spatial
distribution of the SCUSS proper motion catalog in a J2000
equatorial reference frame. The SCUSS proper motion catalog covers
3700 square degrees (Galactic latitude $b < -30^\circ$ and
Equatorial latitude $\delta < -10^\circ$).} \label{fig1}
\end{figure}

The description of reduction of the SCUSS data can be found in Zou
et al.(2014). Photometry was done using SExtractor (Bertin \&
Arnouts 1996). The measured $x$ and $y$ positions were calculated by
applying SExtractor to sources brighter than 1$\sigma$ of the
background level. External astrometric errors were determined by
crossing-matching the SCUSS sources to the UCAC4 sources (Zacharias
et al. 2013). Over the whole survey, the mean offsets of right
ascension and declination between the SCUSS and UCAC4 are
$\triangle\alpha= -0''.001 \pm 0''.008$, $\triangle\delta= -0.''001
\pm 0''.007$, respectively. In this work, we will \textbf{construct}
a proper motion catalog using the measured coordinates ($x$, $y$),
the celestial coordinates ($\alpha$, $\delta$), the $u$-band
magnitudes from SCUSS and the star/galaxy morphological
classifications from SDSS for objects on SCUSS images. The basic
information for each frame including plate size and scale as well as
observation time is also needed in the process of calculation of
proper motions.

\subsection{Astronomic Data from GSC-\uppercase\expandafter{\romannumeral2}}

GSC-\uppercase\expandafter{\romannumeral2} (Lasker et al. 2008), as
first epoch data to obtain the proper motions of SCUSS, was
originally taken from the survey plates observed by the Palomar
Schmidt telescope and the UK Schmidt telescope. Each individual
plate was digitized by the Guide Star Automatic Measuring Machine
(GAMMA) and was processed through image processing and plate
calibration. The final products were stored in the COMPASS database
(Lasker et al. 2008). GSC-\uppercase\expandafter{\romannumeral2}
contains astrometry, multi-band photometry ($R_{F}$, $B_{J}$,
$I_{N}$) and star/non-star classification for objects down to the
limiting magnitude of the survey plates. For stellar sources, the
all-sky average absolute position error per coordinate ranges from
$0.''2$ to $0.''28$ depending on magnitude. Stellar photometry is
determined to $0.13 - 0.22\,\rm{mag}$ as a function of magnitude and
photographic passbands ($R_{F}$, $B_{J}$, $I_{N}$). Outside of the
Galactic plane, the catalogue is complete to $R_{F} = 20
\,\rm{mag}$. There are about 1210 Schmidt plates from different
surveys used in this paper. The main characteristics of the surveys
used in this paper including the epoch, band, limiting magnitude and
sky coverage as well as the number of plates used in this work are
summarized in Table 1. The basic information for objects on each
plate including the measured coordinates, the equatorial
coordinates, magnitude and classification (star/non-star) is needed
in this paper. In addition, The basic information for each plate
from GSC-\uppercase\expandafter{\romannumeral2} data base including
telescope name, plate size and scale, observation time  is also
needed.

\section{CONSTRUCTION OF THE SCUSS PROPER MOTION CATALOG}

Two major aspects determine the final accuracy of absolute proper
motions. One is the measured uncertainties of object positions on
the images of all epochs, and the other one is the systematic error
between different images. The latter one requires unifying images in
different epochs and different bands to one reference system (Qi et
al. 2012).

The general steps of the construction of the SCUSS proper motion
catalog are similar to \textbf{those} of APOP. Relevant equations
and details are recommend to see the paper of Qi et al. (2014). A
few steps are slightly different. The steps are summarized as
follows.

\begin{enumerate}
\item Extract all the
GSC-\uppercase\expandafter{\romannumeral2} plates and SCUSS images
in the same sky field;
\item Compare GSC-\uppercase\expandafter{\romannumeral2}
reference plates with the SDSS, update the star/non-star
classification in GSC with star/galaxy in the SDSS;
\item Reject the synthetic double stars (stars with
the same ID) in each GSC-\uppercase\expandafter{\romannumeral2} plate;
\item Find common objects between the program images and the reference
plate;
\item Smooth out position-dependent difference between each
image pair using stars with small measuring error;
\item Regenerate the $x$, $y$ of
all the objects on program plates using galaxies and remove the MdE,
CdE and residual PdE of proper motions;
\item Calculate the absolute proper motions and new
positions in a given epoch using all the images.
\end{enumerate}

The plates, which are from
GSC-\uppercase\expandafter{\romannumeral2} with good quality and
overlap with the SCUSS, were chosen as reference plates. The
star/galaxy classifications of sources play an important role in the
calculation of the proper motions. However, in
GSC-\uppercase\expandafter{\romannumeral2}, the probability of stars
correctly classified drops with magnitude from better than $90\%$ at
$R_{F} = 18.0\,\rm{mag}$ to less than $50\%$ at $R_{F} =
20.0\,\rm{mag}$, while the percentage that a galaxy is classified as
non-star remains above $90\%$ for all the magnitude range. Outside
of the Galactic plane, stellar classification is reliable to at
least 90\% confidence for magnitudes brighter than $R_{F} =
19.5\,\rm{mag}$ (Lasker et al. 2008). The low accuracy of
star/non-star classification of the fainter objects of the
GSC-\uppercase\expandafter{\romannumeral2} reference plates
eventually affect the error of the proper motions. In our work, most
of the SCUSS survey are covered by the SDSS, star/galaxy separation
of SDSS is about $95\%$ correct at 21.0 mag in $r$ band (Stoughton
et al. 2002). So, we update the
GSC-\uppercase\expandafter{\romannumeral2} star/non-star
classification with the star/galaxy classification from SDSS
morphological classification.

\section{PROPERTIES OF SCUSS PROPER MOTION CATALOG}

The SCUSS proper motion catalog contains the equatorial coordinates
($\alpha$, $\delta$), the absolute proper motions
($\mu_{\alpha}\cos\delta$, $\mu_{\delta}$), and the SCUSS $u$-band
magnitudes for about 7.7 million objects. The proper motions of
quasars cannot be detected by present astrometric observation in the
optical band, so the intrinsic proper motions of quasars can be
considered as zero (Wu et al. 2011). Therefore, the measured proper
motions of quasars can be used to derive the systematic and random
error of the proper motion.

We choose the spectroscopically confirmed quasars from SDSS DR9 (Ahn
et al. 2012). A total of 22831 common quasars among the SDSS and
SCUSS catalog are selected as a quasar sample. The gaussian function
is used to fit the distribution of the quasar proper motions to
derive the systematic and random error. The systematic error of
proper motions is derived from the mean proper motion of the quasar
sample. The dispersion of the proper motions of quasars can be
considered as the random error of proper motions. Figure 2 displays
the histograms of the measured proper motions in right ascension
$\mu_{\alpha}\cos\delta$ (upper) and in declination $\mu_{\delta}$
(lower) for all SDSS spectroscopically confirmed quasars in SCUSS
proper motion catalog. The random errors of the proper motions are
4.2 and $4.4\,\rm{mas/yr}$ in $\mu_{\alpha}\cos\delta$ and
$\mu_{\delta}$, respectively. The systematic error of
$\mu_{\alpha}\cos\delta$ is about $-0.08\,\rm{mas/yr}$ and the
systematic error of $\mu_{\delta}$ is about $-0.06\,\rm{mas/yr}$.
Next, all the 22831 quasars are used to test the position-,
magnitude- and color- dependent errors of the SCUSS proper motions.

\begin{figure}[!htb]
\figurenum{2} \center \resizebox{\width}{!}
{\includegraphics[width=0.9\textwidth]{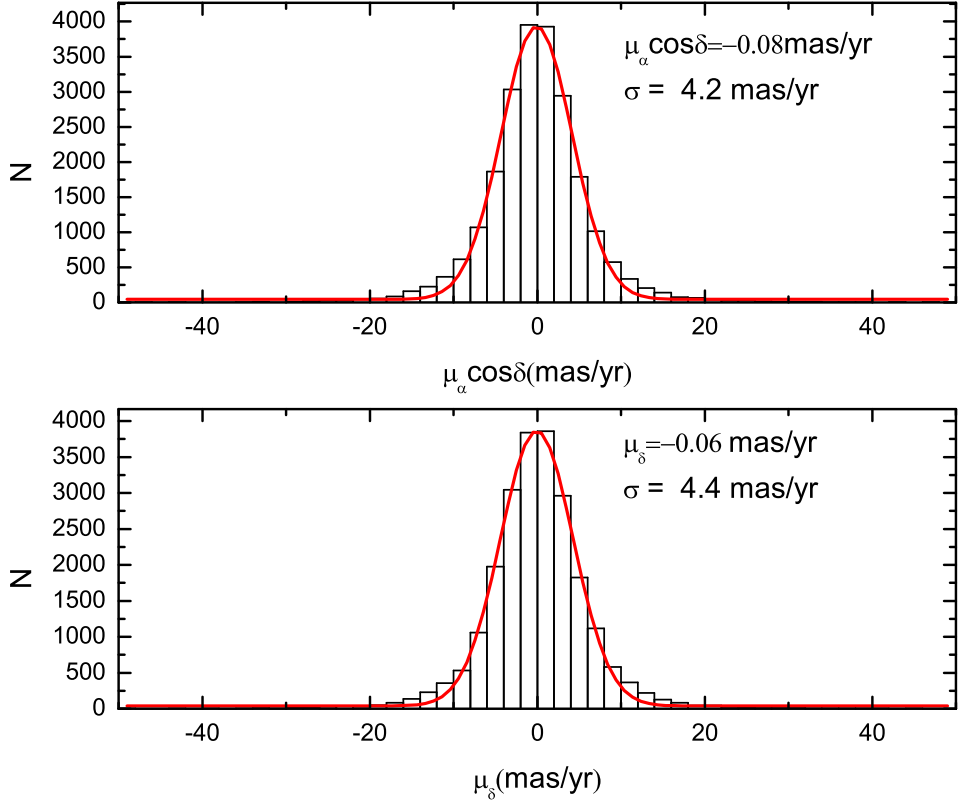}} \caption{The proper
motion distributions of SDSS spectroscopically confirmed quasars in
SCUSS catalog. The upper panel is for $\mu_{\alpha}\cos\delta$ and
the lower panel is for $\mu_{\delta }$. Only quasars with the
absolute values of proper motions less than $50\,\rm{mas/yr}$ are
used for the Gaussian function fitting. The red solid line in each
panel is the best-fitting Gaussian function. The best-fitting
parameters of the Gaussian function including the means and the
dispersions are labeled in each panel.} \label{fig2}
\end{figure}

\subsection{Influence of the Position on the Proper Motion}

The scatters of $\mu_{\alpha}\cos\delta$ (upper) and $\mu_{\delta}$
(lower) of the selected quasar sample as a function of $\alpha$ and
$\delta$ are shown in Figures 3 and 4, respectively. In order to
derive the position-dependent error of the proper motions, quasars
are divided into different position groups with a group size of
$5^\circ$ (in declination) and $10^\circ$ (in right ascension). In
Figures 3 and 4, red points and red error bars show the systematic
and random errors of the quasar proper motions in different groups.
As mentioned above, the systematic and random errors correspond to
the mean and dispersion of the best fit Gaussian function of the
distribution of proper motions for quasar groups.

Figure 3 shows that, for $\mu_{\alpha}$cos$\delta$, there is no
significant dependence of the systematic or random errors in the
$\alpha$ direction. The systematic errors of
$\mu_{\alpha}$cos$\delta$ for quasar groups are small and fluctuate
around zero. The random errors of $\mu_{\alpha}$cos$\delta$ are
about $4\,\rm{mas/yr}$. In addition, there is also no significant
dependence of the errors of $\mu_{\delta}$ in $\alpha$ direction.
The systematic errors of $\mu_{\delta}$ fluctuate around zero. The
random errors of $\mu_{\delta}$ are around  $4\,\rm{mas/yr}$ and do
not significantly change with $\alpha$. From Figure 4, we can see
that there is no significant relation between error (systematic and
random error) of proper motions and $\delta$. The absolute values of
systematic errors of $\mu_{\alpha}$cos$\delta$ and $\mu_{\delta}$
for most quasar groups are less than  $0.1\,\rm{mas/yr}$ and
fluctuate around zero. The random errors of proper motions are about
$4\,\rm{mas/yr}$. So, from Figures 3 and 4, we can find that the PdE
of SCUSS proper motion catalog is well eliminated and the random
errors of the proper motions do not significantly change with
position.

\begin{figure}[!htb]
\figurenum{3} \center \resizebox{\width}{!}{\rotatebox{0}
{\includegraphics[width=0.9\textwidth]{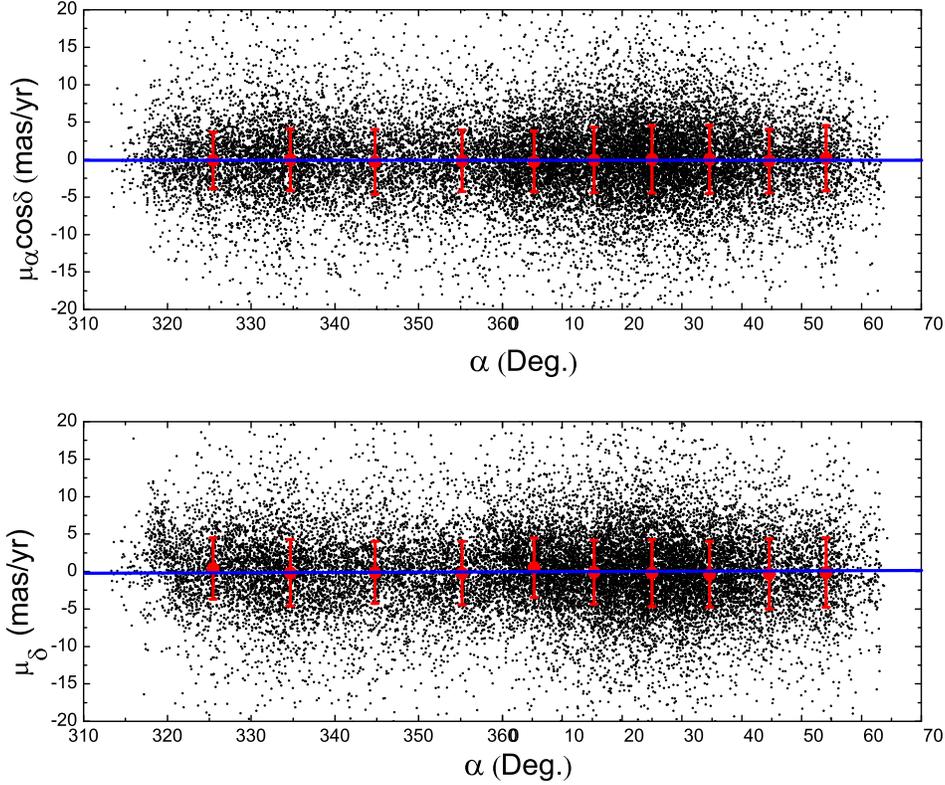}}} \caption{The
scatter of proper motions $\mu_{\alpha}$ cos $\delta$ (upper) and
$\mu_{\delta}$ (lower) for SDSS spectroscopically confirmed quasars
in SCUSS catalog as a function of coordinates in fields located in
the right ascensions from $0^\circ$ to $70^\circ$ and from
$310^\circ$ to $360^\circ$. The systematic and random errors are
showed as red points and red error bars in each panel.} \label{fig3}
\end{figure}

\begin{figure}[!htb]
\figurenum{4} \center \resizebox{\width}{!}{\rotatebox{0}
{\includegraphics[width=0.9\textwidth]{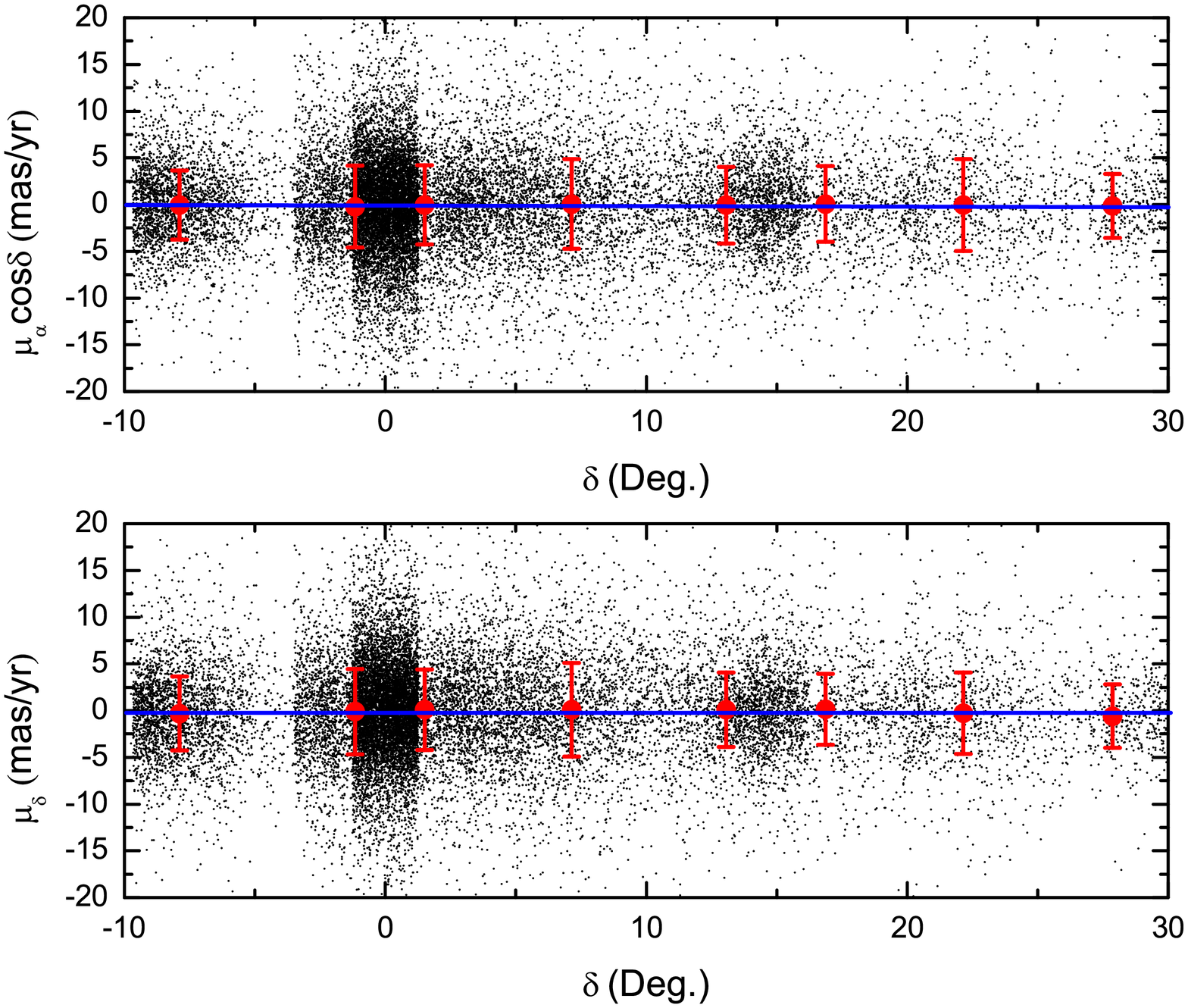}}} \caption{The
scatter of proper motions $\mu_{\alpha}$ cos $\delta$ (upper) and
$\mu_{\delta}$ (lower) for SDSS spectroscopically confirmed quasars
in SCUSS catalog as a function of coordinates in fields located in
the declination from $-10^\circ$ to $30^\circ$.  The systematic and
random errors are showed as red points and red error bars.}
\label{fig4}
\end{figure}

\subsection{Influence of the Magnitude and Color on the Proper Motion}

Figure 5 shows the histogram of SCUSS $u$-band magnitude
distribution of quasar sample. The quasar sample covers the SCUSS
$u$-band magnitude from 16.0 to $24.0\,\rm{mag}$. Quasars are
divided into different magnitude groups with a group size of 1 mag.
Figure 6 shows the scatters of $\mu_{\alpha}$cos$\delta$ (upper) and
$\mu_{\delta}$ (lower) of quasar sample as a function of SCUSS
$u$-band magnitude. In Figure 6, the red points and red error bars
show the systematic and random errors of quasar proper motions in
different groups.

\begin{figure}[!htb]
\figurenum{5} \center \resizebox{\width}{!}{\rotatebox{0}
{\includegraphics[width=0.9\textwidth]{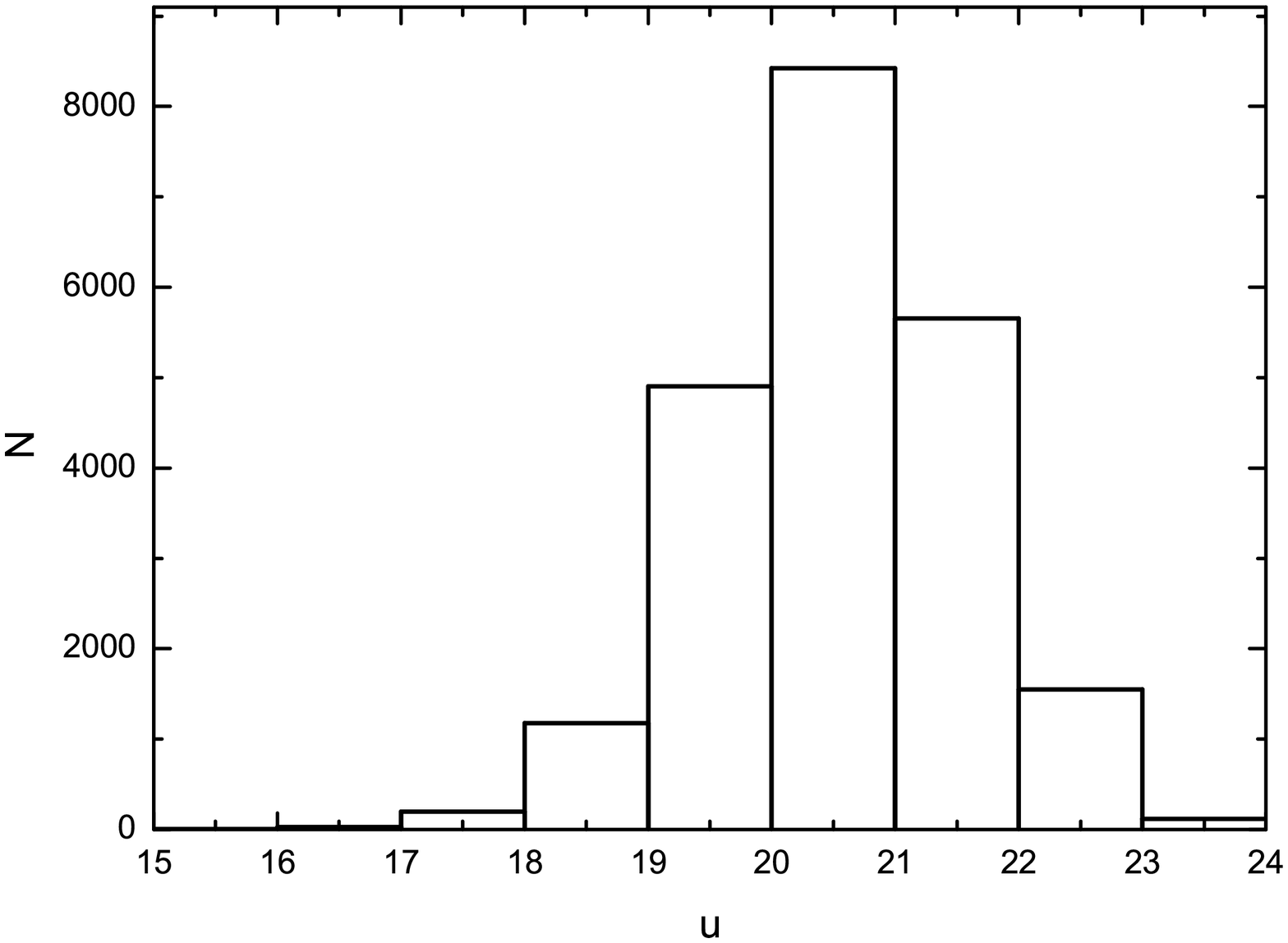}}} \caption{The SCUSS
$u$-band magnitude distribution for SDSS spectroscopically confirmed
quasars in SCUSS catalog.} \label{fig5}
\end{figure}

From Figure 6, we can see that no MdE of the proper motions is
found. The systematic errors of $\mu_{\alpha}$cos$\delta$ and
$\mu_{\delta}$ are both small. The random error of proper motions
increases from the brightest end of the magnitude group to the
faintest end of the magnitude group. For $\mu_{\alpha}$cos$\delta$,
the random error goes up from $2.5\,\rm{mas/yr}$ for quasars
brighter than $u = 18\,\rm{mag}$ to $7.5\,\rm{mas/yr}$ for quasars
fainter than $u = 22\,\rm{mag}$. For $\mu_ {\delta}$, the random
error goes up from $2.6\,\rm{mas/yr}$ for quasars brighter than $u =
18\,\rm{mag}$ to $7.1\,\rm{mas/yr}$ for quasars fainter than $u=
22\,\rm{mag}$. The increasing of the random error may be caused by
the large measuring errors of the fainter sources on image.
$\sigma_\mu = a+b\times \exp ^{(-m/c)}$ is used to describe the
random error of proper motions as a function of magnitude, where
$\sigma_\mu$ is the random error of proper motions, $a$, $b$, and
$c$ are the unknown parameters to be fitted, and $m$ is the SCUSS
$u$-band magnitude. Table 2 lists the best fit parameters for proper
motions in right ascension and declination.

\begin{figure}[!htb]
\figurenum{6} \center \resizebox{\width}{!}{\rotatebox{0}
{\includegraphics[width=0.9\textwidth]{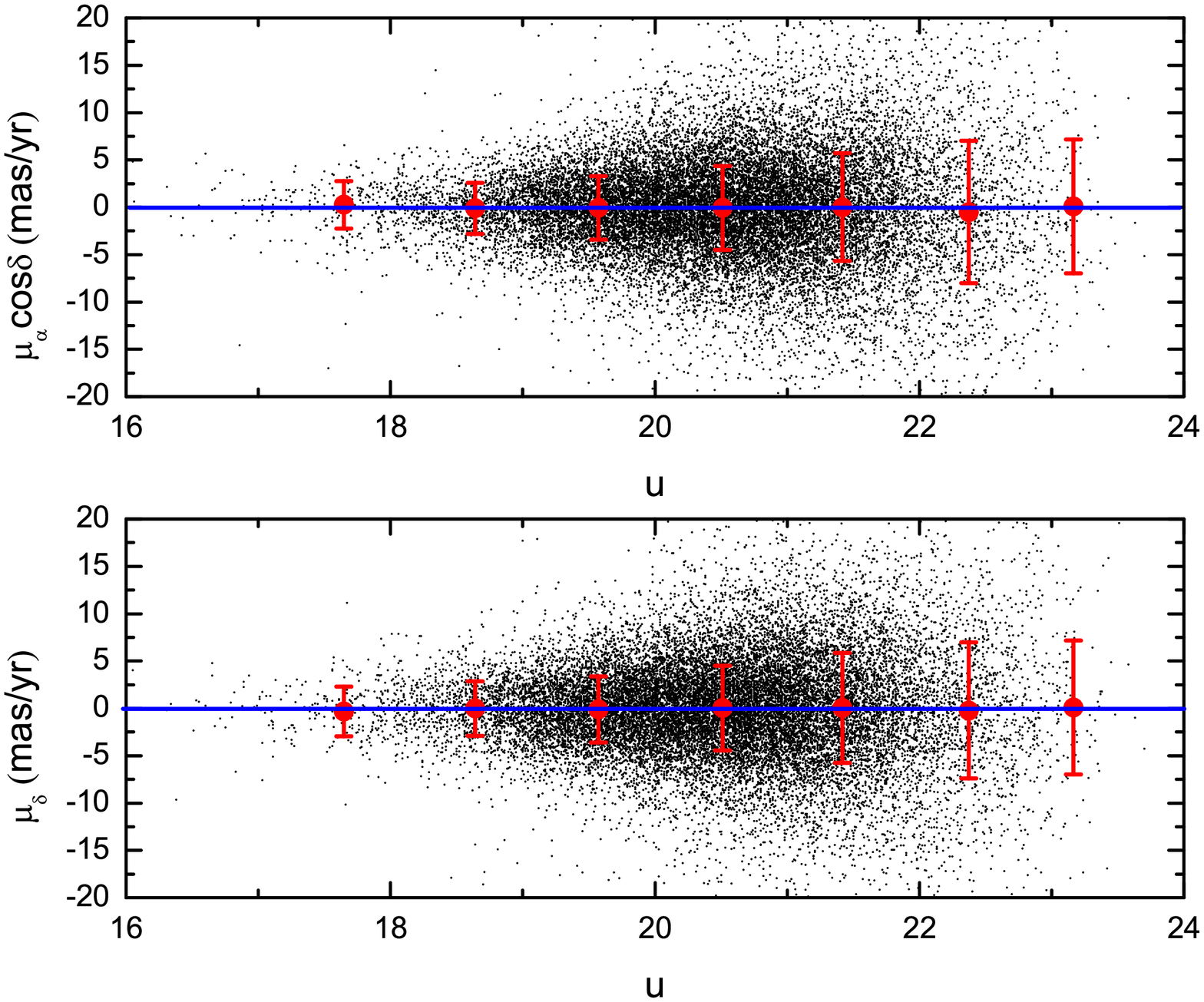}}} \caption{The
scatter of proper motions $\mu_{\alpha}$ cos $\delta$ (upper) and
$\mu_{\delta}$ (lower) for SDSS spectroscopically confirmed quasars
in SCUSS catalog as a function of the SCUSS $u$-band magnitude from
16 to $24\,\rm{mag}$. The systematic and random errors for quasar
groups are showed as red points and red error bars.} \label{fig6}
\end{figure}

The scatters of $\mu_{\alpha}$cos$\delta$ (upper) and $\mu_{\delta}$
(lower) of quasars as a function of SDSS $g-r$ color are shown in
Figure 7. Quasars with $g-r$ color of between $-0.25$ and 1.75 are
divided into different color groups with a group size of 0.3. There
is no relation between error of proper motions and color. The
absolute values of systematic errors of $\mu_{\alpha}$cos$\delta$
and $\mu_ {\delta}$ are around $0.1\,\rm{mas/yr}$. The random errors
of $\mu_{\alpha}$cos$\delta$ and $\mu_{\delta}$ are both about
$4.0\,\rm{mas/yr}$. The CdE of SCUSS proper motion catalog is well
eliminated. The random error of SCUSS proper motion catalog does not
significantly change with color.

\begin{figure}[!htb]
\figurenum{7} \center \resizebox{\width}{!}{\rotatebox{0}
{\includegraphics[width=0.9\textwidth]{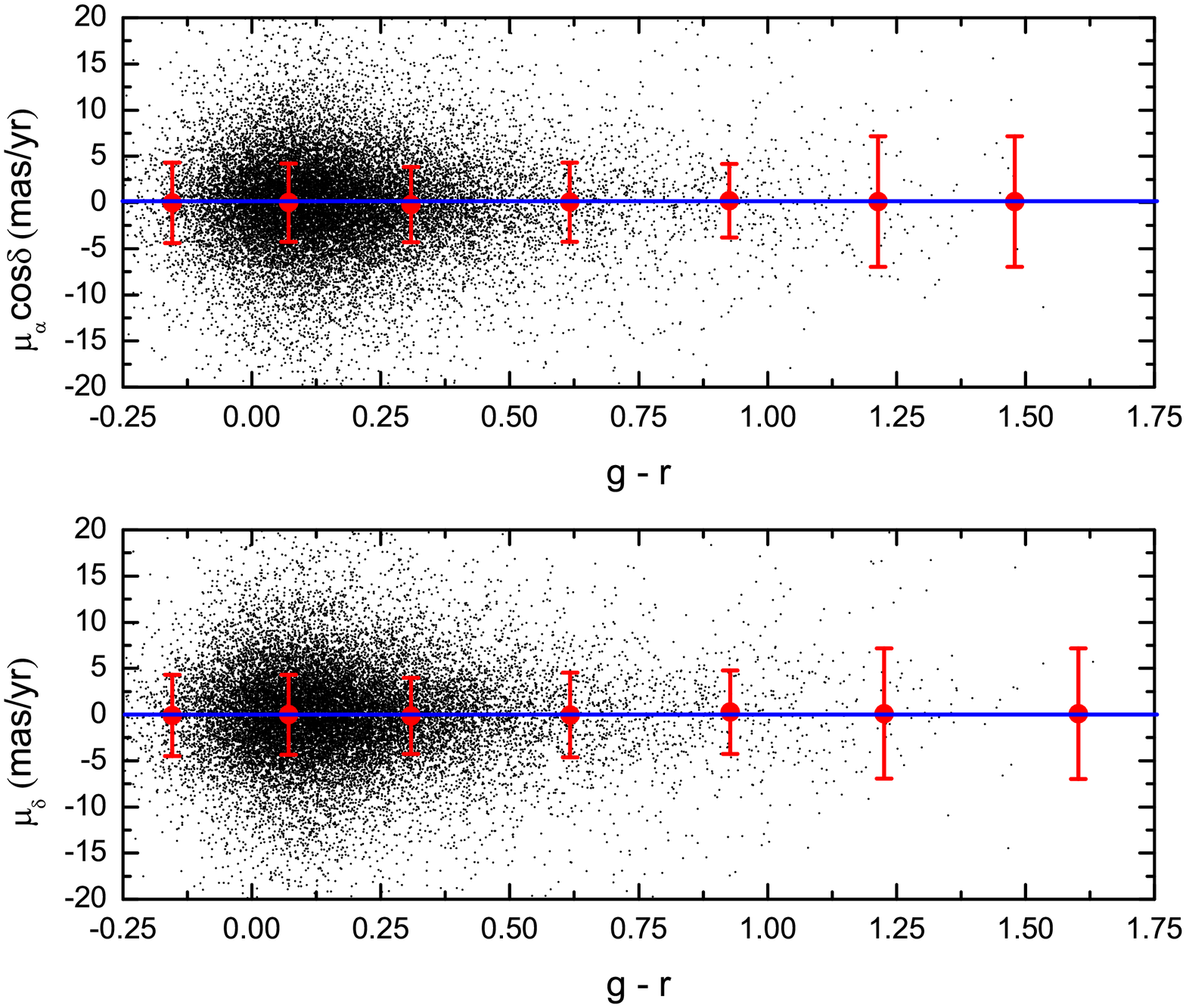}}} \caption{The
scatter of proper motions $\mu_{\alpha}$ cos $\delta$ (upper) and
$\mu_{\delta}$ (lower) for SDSS spectroscopically confirmed quasars
in SCUSS catalog as a function of the SDSS $g-r$ color from $-0.25$
to 1.75. The systematic and random errors for quasar groups are
showed as red points and red error bars.} \label{fig7}

\end{figure}

\section{COMPARISON OF PROPER MOTIONS BETWEEN SCUSS AND SDSS CATALOG}

Comparing the SCUSS proper motion catalog with other catalogs can
estimate the external error of proper motions. The SDSS proper
motion catalog (Munn et al. 2004, 2008; Ahn et al. 2012) covers most
of the SCUSS fields. We compare the proper motions between SCUSS and
SDSS catalog. The SDSS proper motion catalog is derived by combining
SDSS and re-calibrated USNO-B astrometry. For SDSS proper motion
catalog, the random errors in right ascension and declination are
about $4.2\,\rm{mas/yr}$ and $3.7\,\rm{mas/yr}$, respectively (Munn
et al. 2004). The median proper motion for all the quasar sample is
about $0.2\,\rm{mas/yr}$, but the systematic errors can be larger by
a factor of 2 - 3 in small sky patches (Bond et al. 2010; Ahn et al.
2012).

The common stars between SCUSS catalog and SDSS catalog are eslected
and divided into different position ($\alpha$, $\delta$) and
magnitude groups. The means and the dispersions of differences of
proper motions in different common star groups are derived. The mean
differences of proper motions as a function of $\alpha$ are plotted
in Figure 8. The upper panel is the mean differences of
$\mu_{\alpha}$ of stars and the lower panel is the mean differences
of $\mu_{\delta}$ of stars. Figure 8 shows that, for many regions,
the mean differences of proper motions between SDSS and SCUSS are
small. From the upper panel of Figure 8, we can see that, in the
region $60\textordmasculine<\alpha<70\textordmasculine$, the
absolute values of mean differences of $\mu_{\alpha}$ of common
stars are about $1.5\,\rm{mas/yr}$. The lower panel shows that, in
the region $0\textordmasculine <\alpha<10\textordmasculine$,
$60\textordmasculine <\alpha<70\textordmasculine$,
$310\textordmasculine <\alpha<320\textordmasculine$ and
$320\textordmasculine <\alpha<330\textordmasculine$, the absolute
values of mean differences of $\mu_{\delta}$ of common stars are
about  $1.0\,\rm{mas/yr}$. However, in the region
$\alpha=320\textordmasculine$, the absolute value of mean difference
of $\mu_{\delta}$ of stars reaches about $3.5\,\rm{mas/yr}$. In most
regions, the mean differences of the proper motions are smaller than
the random errors of SDSS and SCUSS proper motions. The means and
dispersions of differences of proper motions as a function of
$\delta$ are plotted in Figure 9. From Figure 9, we can see that the
mean differences of proper motions between SDSS and SCUSS are small.
The absolute values of mean differences are less than
$0.5\,\rm{mas/yr}$.

The means and dispersions of differences of proper motions as a
function of SCUSS $u$-band magnitude are plotted in Figure 10.
Figure 10 shows that, in the magnitude range $u > 17\,\rm{mag}$, the
mean differences of proper motions between SDSS and SCUSS are small
and the dispersion of the differences of proper motions increases
with magnitude.

\begin{figure}[!htb]
\figurenum{8} \center \resizebox{\width}{!}{\rotatebox{0}
{\includegraphics[width=0.9\textwidth]{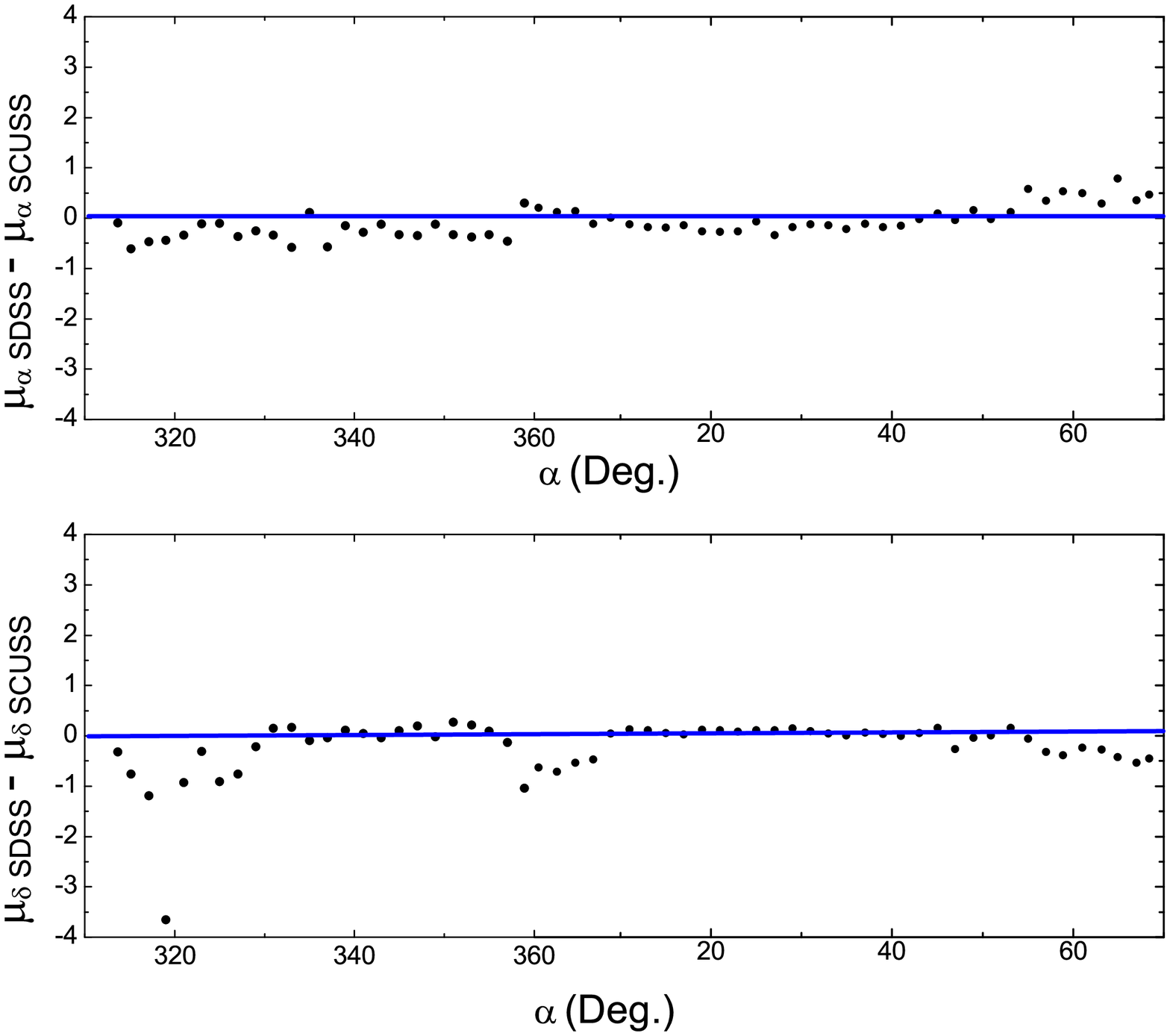}}} \caption{ The mean
differences of proper motions $\mu_{\alpha}$ (upper) and
$\mu_{\delta}$ (lower) of common stars between SDSS and SCUSS as a
function of $\alpha$. The typical dispersion of differences of
proper motions is about $5\,\rm{mas/yr}$.} \label{fig8}
\end{figure}

\begin{figure}[!htb]
\figurenum{9} \center \resizebox{\width}{!}{\rotatebox{0}
{\includegraphics[width=0.9\textwidth]{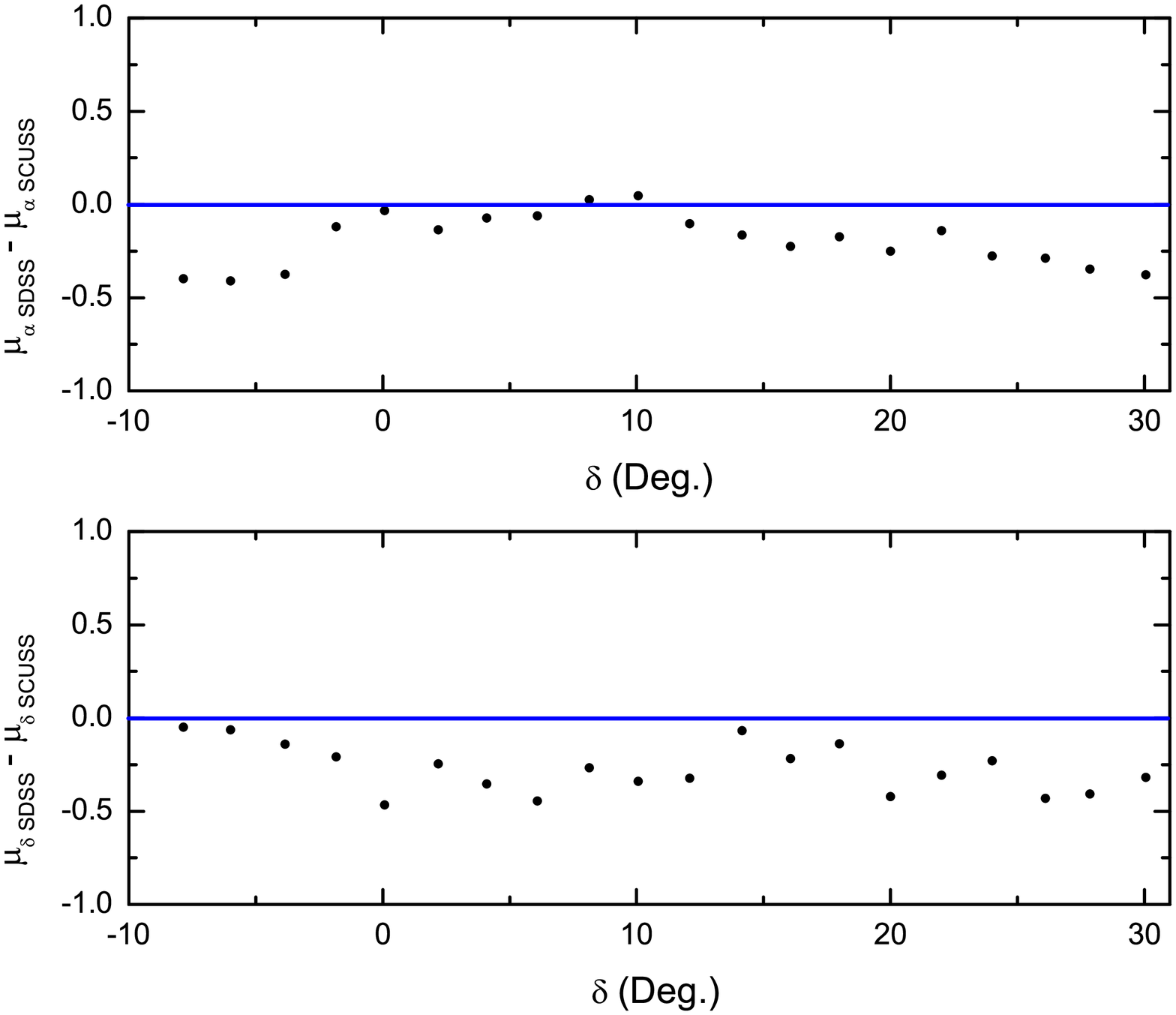}}} \caption{ The mean
differences of proper motions $\mu_{\alpha}$ (upper) and
$\mu_{\delta}$ (lower) of common stars between SDSS and SCUSS as a
function of $\delta$. The typical dispersion of differences of
proper motions is about $5\,\rm{mas/yr}$.} \label{fig9}
\end{figure}

 \begin{figure}[!htb]
\figurenum{10} \center \resizebox{\width}{!}{\rotatebox{0}
{\includegraphics[width=0.9\textwidth]{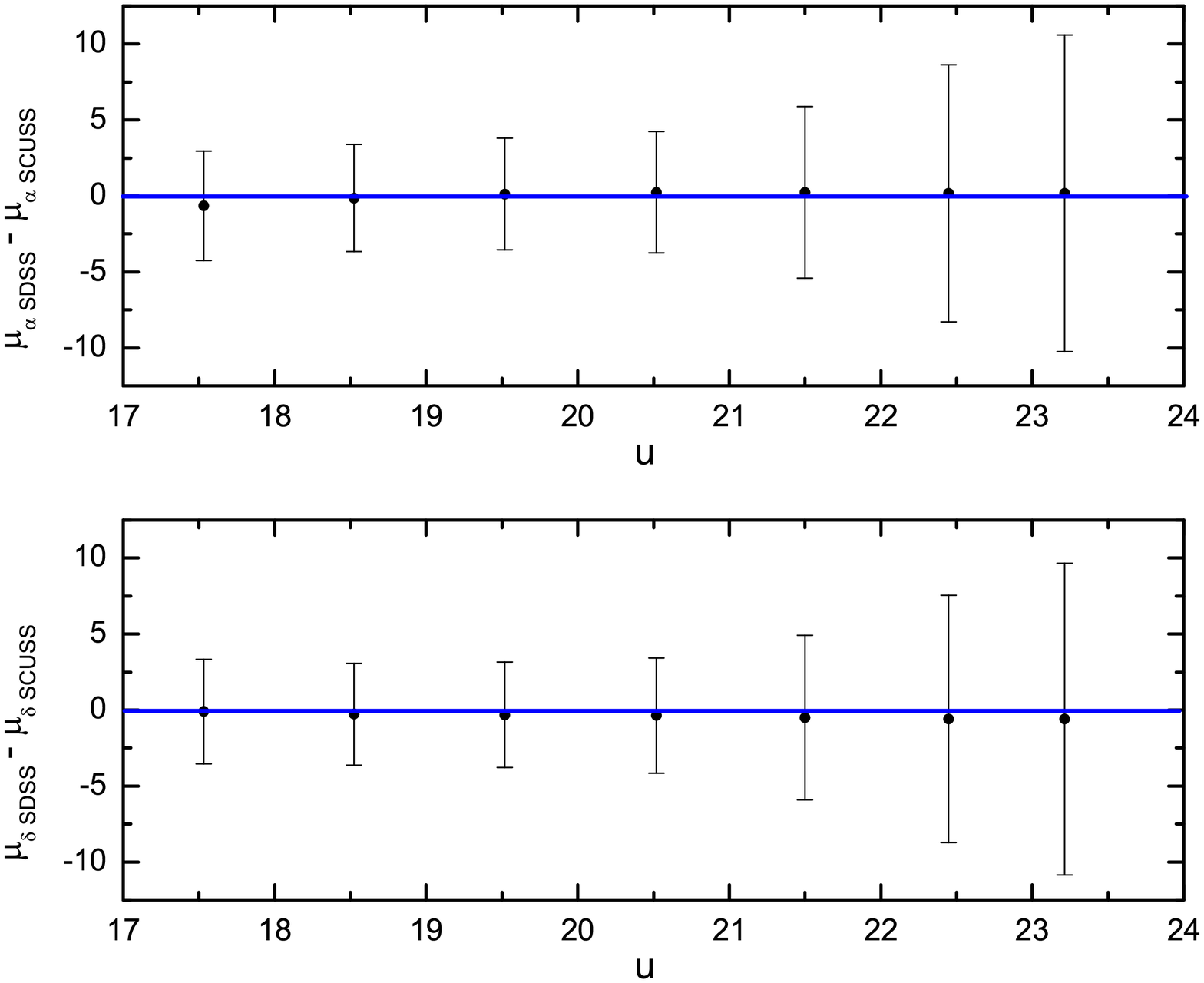}}} \caption{The means
and dispersions of differences of proper motions $\mu_{\alpha}$
(upper) and $\mu_{\delta}$ (lower) of common stars between SDSS and
SCUSS as a function of the SCUSS $u$-band magnitude.} \label{fig10}
\end{figure}

\section{CONCLUSION}

The absolute proper motions of SCUSS program are computed by
combining the SCUSS and GSC-\uppercase\expandafter{\romannumeral2}.
The basic method of calculation of absolute proper motions follows
that of APOP. The measured coordinates ($x$, $y$) of objects instead
of equatorial coordinates are used to calculate the absolute proper
motions. The stellar sources with good imaging quality are used to
eliminate the PdE between the reference and program image. The
galaxies with good imaging quality are used to eliminate the MdE,
CdE and the residual PdE.

The 22831 common quasars among SCUSS and SDSS are selected to test
the properties of SCUSS proper motion catalog. The systematic errors
of the proper motions are $-0.08$ and $-0.06\,\rm{mas/yr}$ in
$\mu_{\alpha}$cos$\delta$ and $\mu_{\delta}$. In addition, the
random errors of the proper motions are 4.2 and $4.4\,\rm{mas/yr}$
in $\mu_{\alpha}$cos$\delta$ and $\mu_{\delta}$. We made a detailed
analysis of the obtained proper motions in order to determine the
position-, magnitude- and color- dependent errors of the proper
motions. Finally, we find that the PdE, MdE and CdE of SCUSS proper
motion catalog are small. The position- and color- dependent random
errors remain about the same. The random errors of
$\mu_{\alpha}$cos$\delta$ and $\mu_{\delta}$ increase with
magnitude. Beyond that, no significant relations between errors
(systematic and random) and the position, magnitude and color are
found.

The common stars between SCUSS and SDSS catalog are selected. The
means and dispersions of differences of proper motions between SCUSS
and SDSS are derived. The mean differences of proper motions between
SDSS and SCUSS as a function of $\alpha$ are small except for some
regions. There is no relation between the mean differences of proper
motions and $\delta$. The mean differences of proper motions are
small in magnitude region $u>17\,\rm{mag}$.

The SCUSS proper motion catalog with small systematic error and
low random error can be used for many astronomical studies such as
the Galaxy disk kinematics, substructure of the Galaxy and Galaxy
rotation curve.

The release of the SCUSS proper motion catalog is in preparation.
The final version of the SCUSS proper motion catalog will be
available when the SCUSS data becomes public. If you need a
preliminary version of SCUSS proper motion catalog, you can email
Xiyan Peng (xypeng@bao.ac.cn).

\section*{ACKNOWLEDGMENTS}

We would like to thank the anonymous referee for providing rapid and thoughtful report that helped improve the original manuscript greatly.
This work is supported by National Basic Research Program of China (973 Program),
No.2014CB845702, No. 2014CB845704, and No. 2013CB834902.
This work is also supported by the National Natural Science Foundation of
China (NSFC, Nos.11433005, 11373035, 11373033, 11373003, 11273003,
11203034, 11203031,  11303038, 11303043). The project
is managed by the National Astronomical Observatory of China and
Shanghai Astronomical Observatory. This work is also supported by
the joint fund of Astronomy of the National Nature Science
Foundation of China and the Chinese Academy of Science, under Grants
U1231113. The SCUSS is funded by the Main Direction Program of
Knowledge Innovation of Chinese Academy of Sciences (No.
KJCX2-EW-T06). It is also an international cooperative project
between National Astronomical Observatories, Chinese Academy of
Sciences and Steward Observatory, University of Arizona, USA.
Technical supports and observational assistances of the Bok
telescope are provided by Steward Observatory.

\clearpage


\begin{table}
\caption{Plate Information from
GSC-\uppercase\expandafter{\romannumeral2} }
\begin{tabular}{llcccc}
\hline  Survey   & Epoch  & Band & Depth & plates used & sky
coverage \\
\hline
 {Pal-QV} & 1983-85  & $V_{12}$ &19.5 &103 & $\delta \geq 0^{\circ }$  \\
 SERC-EJ & 1979-88  & $B_{J}$ & 23.0 &72 & $ -15 < \delta \leq  0$   \\
 POSS-$\uppercase\expandafter{\romannumeral1}$ E & 1950-1958   & $E$ & 20.0 &144 & $ \delta \geq  -30$    \\
 POSS-$\uppercase\expandafter{\romannumeral2}$ O & 1950-1958  & $O$ & 21.0 &143 & $ \delta \geq  -30$      \\
 POSS-$\uppercase\expandafter{\romannumeral2}$ J & 1987-00  & $B_{J}$ & 22.5 &141 & $ \delta \geq  0$    \\
 POSS-$\uppercase\expandafter{\romannumeral2}$ F & 1987-99  & $R_{F}$ & 20.8 &141 & $ \delta \geq  0$     \\
 POSS-$\uppercase\expandafter{\romannumeral2}$ N & 1989-02  & $I_{N}$ & 19.5 &139 & $ \delta \geq  0$    \\
 SERC ER & 1990-98  &  $R_{F}$ & 22.0 &93 & $ -15 < \delta \leq  0$   \\
 SERC I & 1990-02  & $I_{N}$ & 19.5 &93 & $ \delta \leq  0$   \\
\hline
\end{tabular}
\end{table}


\begin{table}
\begin{minipage}{120mm}
\caption{Best-Fitting Parameters to Describe the Magnitude
Dependence of the Random Errors of Proper Motions}\label{t2.tab}
\begin{tabular}{cccc}
\hline Type of $\sigma_\mu$ & a   & b  & c \\
\hline
$\mu_{\alpha}$ cos $\delta$  & $-4.78$  & 1.08  & $-9.47$\\
$\mu_{\delta}$  & $-4.77$  & 1.23  & $-10.07$    \\

\hline
\end{tabular}
\end{minipage}
\end{table}

\clearpage

\end{document}